\documentclass[aps,prb,twocolumn,superscriptaddress,showpacs,floatfix]{revtex4}
\usepackage{graphicx}

\begin{document}

\title{Effects of nanoscale spatial inhomogeneity in strongly correlated 
systems}

\author{M. F. Silva}
\affiliation{Instituto de F\'{\i}sica,
Universidade de S\~ao Paulo,
Caixa Postal 66318, 05315-970 S\~ao Paulo, SP, Brazil}

\author{N. A. Lima}
\affiliation{Colegiado de Engenharia de Produ\c{c}\~ao, 
Funda\c{c}\~ao Universidade Federal do Vale do S\~ao Francisco, 
Caixa Postal 252, 56306-410 Petrolina, PE, Brazil}

\author{A. L. Malvezzi}
\affiliation{Departamento de F\'{\i}sica,
Faculdade de Ci\^encias,
Universidade Estadual Paulista,
Caixa Postal 473, 17015-970 Bauru, SP, Brazil} 

\author{K. Capelle}
\email{capelle@if.sc.usp.br}
\affiliation{Departamento de F\'{\i}sica e Inform\'atica,
Instituto de F\'{\i}sica de S\~ao Carlos,
Universidade de S\~ao Paulo,
Caixa Postal 369, 13560-970 S\~ao Carlos, SP, Brazil}
\date{\today}

\begin{abstract}
We calculate ground-state energies and density distributions of Hubbard
superlattices characterized by periodic modulations of the on-site 
interaction and the on-site potential. Both density-matrix renormalization
group and density-functional methods are employed and compared. We find
that small variations in the on-site potential $v_i$ can simulate, cancel, 
or even overcompensate effects due to much larger variations in the on-site 
interaction $U_i$. Our findings highlight the importance of nanoscale spatial
inhomogeneity in strongly correlated systems, and call for reexamination 
of model calculations assuming spatial homogeneity.
\end{abstract}

\pacs{pacs}



\maketitle

\newcommand{\be}{\begin{equation}}
\newcommand{\ee}{\end{equation}}
\newcommand{\bea}{\begin{eqnarray}}
\newcommand{\eea}{\end{eqnarray}}
\newcommand{\bi}{\bibitem}

\renewcommand{\r}{({\bf r})}
\newcommand{\rp}{({\bf r'})}

\newcommand{\ua}{\uparrow}
\newcommand{\da}{\downarrow}
\newcommand{\la}{\langle}
\newcommand{\ra}{\rangle}
\newcommand{\dg}{\dagger}

A large part of the complexity of strongly correlated systems arises from
the multiple phases that coexist or compete in their phase diagrams. 
Metallic and insulating phases are separated by metal-insulator transitions,
and subject to the formation of various types of long-range order, such as
antiferromagnetism, superconductivity, and charge or spin-density waves.
The relative stability of such phases is determined by differences in
appropriate thermodynamic potentials, or, at zero temperature, in their
ground-state energies. Identification of the appropriate order parameters 
and calculation of the ground-state energies of the various phases is a 
complicated problem, and the nature of the phase diagram of many strongly
correlated systems is still subject to considerable controversy. It is
widely believed, however, that a minimal model containing the essence of
strong correlations, and displaying many of the above-mentioned phases,
is the homogeneous Hubbard model, which in one dimension and standard 
notation reads
\be
\hat{H}_{hom}=
-t\sum_{i,\sigma} (c_{i\sigma}^\dagger c_{i+1,\sigma}+H.c.)
+U\sum_i c_{i\ua}^\dagger c_{i\ua}c_{i\da}^\dagger c_{i\da}.
\label{hm}
\ee
Much theoretical effort is thus going into the analysis of the homogeneous
Hubbard model and the clarification of the nature of its ground state.

In a parallel development, nanoscale spatial inhomogeneity has been observed
experimentally to be a ubiquitious feature of strongly correlated 
systems,\cite{inhom1,inhom2,inhom3,inhom4,inhom5,inhom6,inhom7,inhom8} but 
although its importance is widely recognized, the consequences of such 
inhomogeneity are still insufficiently understood. The present paper 
investigates the effects of, and the competition between, two different 
manifestations of nanoscale inhomogeneity in strongly correlated systems: 
local variations in the on-site potential and in the on-site interaction. 
We base our analysis on the {\it inhomogeneous} Hubbard model
\bea
\hat{H}_{inhom}=
-t\sum_{i,\sigma} (c_{i\sigma}^\dagger c_{i+1,\sigma} + H.c.)
\nonumber \\
+\sum_i U_i c_{i\ua}^\dagger c_{i\ua}c_{i\da}^\dagger c_{i\da}
+\sum_{i\sigma} v_i c_{i\sigma}^\dagger c_{i\sigma},
\label{ihm}
\eea
which differs from the homogeneous model (\ref{hm}) by allowing for spatial 
variations in the on-site interaction $U_i$ and the presence of the on-site 
potential $v_i$. Variations in $U_i$ and $v_i$ may arise, e.g.,
due to inequivalent sites in the natural unit cell, modulation of system 
parameters in artificial heterostructures, or self-consistent modulations 
in local system properties due to formation of charge-ordered states.
In this paper we are specifically concerned with one-dimensional superlattice 
structures in which both $U_i$ and $v_i$ vary periodically on a length 
scale comparable to, or somewhat larger than, the lattice constant. 
Such superlattices have recently attracted much attention due to their 
complex ground-state and transport properties.\cite{superl1,superl2,superl3,superl4,superl5,superl6,superl7,superl8,superl9,superl10}
Our results, reported below, have a direct bearing on the investigation of
such superlattices. However, for our present purposes the most important
aspect of superlattice structures is that they constitute a representative 
system in which the consequences of nanoscale spatial variations of system 
parameters in the presence of strong Coulomb correlations can be explored 
systematically. Accordingly, we expect our main conclusions to hold also
in many other spatially inhomogeneous correlated systems.

\begin{table}
\caption{\label{table1}
Ground-state energy of two open superlattices with modulated on-site
interaction $U_i$ and spatially constant on-site potential $v_i$, obtained
with DMRG and with DFT/BA-LDA.
Upper part: large lattice with $L=300$ sites. $L_U=10$ interacting sites
($U_i=3$) alternate with $L_0=10$ noninteracting sites ($U_i=0$).
Lower part: strongly modulated lattice with $L=100$ sites. $L_U=1$
interacting site ($U_i=6$) alternates with $L_0=1$ noninteracting site.
$N$ is the number of fermions, and the column labeled $\Delta \%$ contains
the absolute percentual deviation of the DMRG from the DFT/BA-LDA values.
The agreement between DMRG and BA-LDA is slightly better for the more
slowly modulated lattice.\cite{footnote1}}
\begin{ruledtabular}
\begin{tabular}{c|c|c|c}
$N$ & $E_0^{\rm DMRG}/t$ & $E_0^{\rm BA-LDA}/t$ & $\Delta \% $ \\
\hline
50  &-97.994&-98.342&0.35\\
100 &-185.66&-187.32&0.89\\
150 &-255.62&-258.65&1.17\\
200 &-302.10&-305.68&1.17\\
250 &-321.09&-324.02&0.90\\
300 &-310.35&-311.87&0.49\\
\hline
40  &-69.822&-71.110&1.81\\
50  &-82.078&-83.730&1.97\\
75  &-100.14&-103.11&2.88\\
80  &-101.81&-104.94&2.98\\
120 &-79.947&-79.323&0.79
\end{tabular}
\end{ruledtabular}
\end{table}

\begin{figure}
\includegraphics[height=80mm,width=50mm,angle=-90]{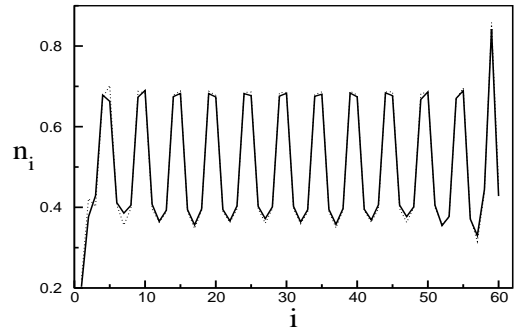}
\caption{\label{fig1} Density profile of a one-dimensional superlattice
with $L=60$ sites, $N=30$ fermions, open boundary conditions, and a
superlattice structure consisting of a periodic sequence of $L_U=3$
interacting ($U_i=3$) and $L_0=2$ noninteracting ($U_i=0$) sites. Full
curve: DMRG calculation. Dotted curve: DFT/BA-LDA calculation.}
\end{figure}

Figure~\ref{fig1} shows the density profile of a typical superlattice
structure in which the on-site interaction $U_i$ is modulated in a repeated
pattern of repulsive ($U_i=3$) and noninteracting ($U_i=0$) 'layers' with
$L_U$ and
$L_0$ sites, respectively, and the on-site potential $v_i$ is taken to
be constant at all sites. The two curves shown were obtained with different
many-body techniques. The full curve was obtained using the density-matrix
renormalization group (DMRG),\cite{dmrg1,dmrg2} while the
dotted curve was obtained from density-functional theory (DFT) within the
Bethe-Ansatz local-density approximation (BA-LDA).\cite{balda1,balda2,balda3}
In view of the complexity of the problem and the surprising nature of some
of our conclusions, we found it advisable to bring two independently
developed and implemented many-body methods to bear on the problem.

DMRG is a well-established numerical technique, whose precision can be
improved systematically, at the expense of increased computational 
effort.\cite{dmrg1,dmrg2} In our DMRG calculations, truncation errors were 
kept of the order of $10^{-6}$ or smaller, and increasing the precision beyond 
this did not affect any of our conclusions. BA-LDA is a more recent 
development\cite{balda1,balda2,balda3}
(although the original LDA concept is, of course, widely used in {\it ab
initio} calculations). In LDA calculations the final precision is ultimately
limited by the locality assumption inherent in the LDA, and improvements must
come from the development of better functionals. This intrinsic
limitation of LDA is offset by its applicability to very large and
inhomogeneous systems, at much reduced computational effort: Calculations
for the type of superlattice structures investigated here typically take
only seconds to minutes with BA-LDA, regardless of the type of boundary
condition used.\cite{footnote0} Final BA-LDA results for densities and
energies typically agree with DMRG ones to within $\alt 3\%$, the agreement
being slightly better for energies than for densities.\cite{footnote1}
Here we consider both
methods as complementary. All essential conclusions reported below were
obtained on the basis of independently implemented and performed BA-LDA and
DMRG calculations. As an illustration, Table \ref{table1} compares ground-state
energies obtained with both methods for one much larger and one much more
rapidly modulated superlattice than the one shown in Fig.~\ref{fig1}.

\begin{figure}
\includegraphics[height=95mm,width=60mm,angle=-90]{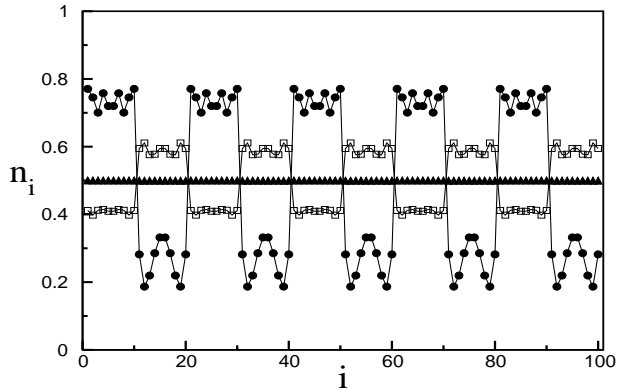}
\caption{\label{fig2} Density profiles, obtained with BA-LDA,\cite{footnote2}
of an $L=100$ site system of $N=50$ fermions with
periodically modulated on-site interaction of amplitude $U=3$ ($L_U=L_0=10$),
and periodic boundary conditions. Squares: no modulation in on-site
potential ($v_i=0$). Circles: on-site potential modulated such that
$v_i=-2$ on the interacting sites and $v_i=0$ on the noninteracting sites.
The density profile is inverted, indicating overcompensation of $U_i$ by $v_i$.
Triangles: on-site potential modulated such that $v_i=-0.555$ on the
interacting sites and $v_i=0$ on the noninteracting sites. The superlattice
structure is erased from the density profile.
The lines are guides for the eye.}
\end{figure}

Inspection of Fig.~\ref{fig2} shows that an attractive potential on the
repulsively interacting sites can completely reverse the effect of the Coulomb
repulsion $U_i$ and draw a substantial number of electrons to the interacting
sites (circles in Fig.~\ref{fig2}). While this might have been anticipated
qualitatively as a result of the competition between an attraction and a
repulsion, it comes as a surprise that the effect of the (often neglected)
variations in the on-site potential is much stronger than the one of variations
in the on-site interaction: already a very weak attractive potential suffices
to smooth out the density distribution, resulting in an essentially homogeneous
charge profile (triangles in Fig.~\ref{fig2}). Although we have taken
superlattices as our example, the effect is clearly not dependent on
periodicity of the modulations in $U_i$ and $v_i$, and is expected to
show up rather generally.
                                                                                
Figure \ref{fig1} and Table \ref{table1} represent superlattices in which only
the on-site interaction is spatially modulated, which is the case mostly
studied in the literature.\cite{superl1,superl2,superl3,superl4,superl5,superl6,superl7,superl8,superl9,superl1} Many important
features of superlattice structures are already apparent in this type of
model. However, in a real system it is impossible to modulate the on-site
interaction without simultaneously modulating the on-site potential as well,
i.e., without creating inequivalent sites. Such a double modulation is found,
e.g., in artificially grown layered structures, in impurity systems, and in
periodic arrays of Fermi-liquid leads (corresponding to approximately
noninteracting sites) and quantum wires/dots (corresponding to interacting
sites). Fig.~\ref{fig2} illustrates the consequences a modulation of the
on-site potential has on the density profile of a system in which both
interaction and potential vary.

\begin{figure}
\includegraphics[height=80mm,width=50mm,angle=-90]{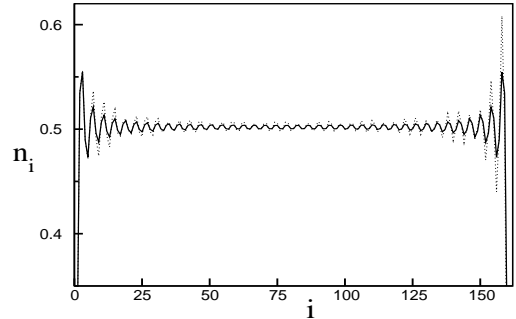}
\caption{\label{fig3} Full curve: density profile, obtained with
BA-LDA,\cite{footnote2} of a $L=160$ site
homogeneous system with open boundary conditions, $U_i=2$, $N=80$, $v_i=0$.
Friedel oscillations arising from the system boundaries are clearly visible.
Dotted curve: density profile of same system but subject to modulations
periodically alternating $L_U=6$ sites with $U_i=2$ and $v_i=-0.2$, with
$L_0=10$ sites with $U_i=v_i=0$. The superlattice structure due to the
presence of both modulations is completely erased from the density profile,
while the Friedel oscillations arising from the boundary remain prominent.}
\end{figure}

In Fig.~\ref{fig3} we compare the Friedel oscillations arising from the system
boundaries in a homogeneous system with the ones arising in a superlattice
of same size and with the same number of fermions, but subject to periodic
modulations of $U_i$ and $v_i$, chosen such that both density profiles become
similar. We have deliberately not chosen modulation parameters that optimize
the agreement between both curves, because had we done so they would be
visually indistinguishable on this scale.

Figure~\ref{fig2} shows that a small value of $v_i$ can have stronger effects 
than a larger value of $U_i$, while Fig.~\ref{fig3} shows that essentially the 
same modulation pattern in the density profile can be obtained from either $v$ 
or $U$. These are unexpected findings. Normally it is assumed that in systems 
modeled by the Hubbard model the particle-particle interaction $U$ is much 
more important than the on-site potential, which is mostly taken to be 
spatially constant, or, if it varies, to produce only minor modifications 
in systems whose physics is governed by $U$.

\begin{figure}
\includegraphics[height=80mm,width=50mm,angle=-90]{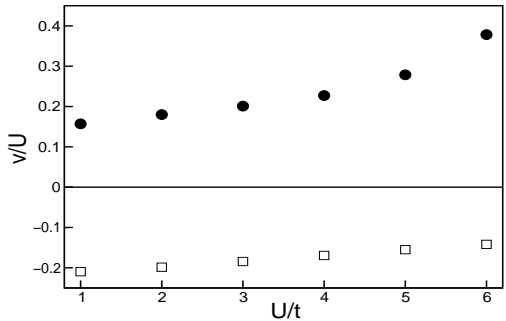}
\caption{\label{fig4}
Open squares: Amplitude of the modulation of the attractive on-site potential
that cancels as much as possible the effect of the modulation of the repulsive
on-site interaction, leading to a homogeneous density profile [criterium (i)].
Full circles: Amplitude of the modulation of the attractive on-site potential
that reproduces in the doubly modulated system the energy of the
homogeneous system [criterium (ii)].\cite{footnote3}
System parameters: $L=160$ sites, $N=80$ fermions, $L_U=6$ interacting sites,
alternating with $L_0=10$ noninteracting sites.}
\end{figure}

To investigate in more detail this competition between on-site interaction 
and on-site potential we need to establish criteria for comparing the 
consequences of $v_i$ and of $U_i$. Motivated by the experimental observation 
of nanoscale density variations and by the theoretical importance of 
ground-state energies for analyses of phase diagrams, we adopt two distinct
criteria. Criterium (i) consists in searching for that modulation of the 
on-site potential $v_i$ in a doubly modulated lattice that cancels the effect 
of the modulation of $U_i$ on the density distribution, i.e., smoothes out 
the oscillations, making the net density homogeneous. 
A particular example of this cancellation is given by the triangles in 
Fig.~\ref{fig2}. Criterium (ii) consists in searching for that modulation of 
the on-site potential $v_i$ in the doubly modulated lattice that yields the 
same ground-state energy $E_0$ as in a homogeneous lattice with $v_i=0$ and
$U_i=U$ at all sites.\cite{footnote3} Our results, displayed in 
Fig.~\ref{fig4},\cite{footnote0} show that, regardless of whether one adopts 
the density or the energy criterium, {\em the modulation of the on-site 
potential required to cancel the effect of the modulation of the on-site 
interaction is up to an order of magnitude smaller than $U$}. For open
boundary conditions we have obtained the same conclusion also from DMRG 
calculations. Changes in the modulation pattern do not change the order
of magnitude of the ratio of $|v|$ to $U$ appreciably.

A semi-quantitative explanation for this relation of $|v|$ to $U$ can be given 
within DFT, by considering the effective potential entering the Kohn-Sham 
equations for the Hubbard model, $v_{eff,i}=v_{ext,i}+v_{H,i}+v_{c,i}$. For
unpolarized systems ($n_{\uparrow,i} = n_{\downarrow,i}=n_i/2$) the Hartree
potential $v_{H,i}$ can be written $v_{H,i}= U_i n_i/2$. Within BA-LDA DFT the 
density and total energy are thus calculated from an effective Hamiltonian 
containing the modulated interaction and external potential only via the 
combination $v_{ext,i}+U_i n_i/2 + v_{c,i}(n_i,U_i)$. Since
the correlation potential $v_{c,i}$ is typically about an order of magnitude 
smaller than $v_{H,i}$, the modulated interaction $U_i$ enters the effective
Hamiltonian approximately on the same footing as the modulated potential 
$v_{ext,i}$, but renormalized by the factor $n_i/2$. In the above calculations 
the average density $N/L =0.5$, and the local density $n_i$ is not very
different. The upshot is that self-consistent screening of the particle-particle
interaction effectively reduces the modulation in the interaction by a factor 
$\sim 4$, compared to modulations in the potential, in good agreement with the 
numerical results in Fig.~\ref{fig4}.\cite{thanks}

Of course, in a real system one cannot adjust $v_i$ at will, and the precise
fine-tuning required to obtain smooth density profiles, or energies identical 
to the ones found in homogeneous systems, is not expected to occur frequently 
in nature. The main implication of these criteria is rather that they establish
a scale for comparison of $v$ and $U$, indicating that even weak spatial 
variations of $v$ can be more important than much stronger ones in $U$. This 
observation flags a warning signal to the use of homogeneous Hubbard models 
(or ones in which only $U$ is modulated) in the analysis of situations 
characterized by nanoscale spatial inhomogeneity, such as the pseudo-gap
phase of cuprates\cite{inhom1,inhom2,inhom3,inhom4,inhom5,inhom6,inhom7,inhom8}
or superlattices and similar heterostructures.\cite{superl1,superl2,superl3,superl4,superl5,superl6,superl7,superl8,superl9,superl10}

We conclude that even in the presence of strong correlations, spatial 
variations of the on-site potential $v_i$ are not a minor complication in a 
system dominated by the on-site interaction $U_i$, but a major effect, which 
crucially contributes to observables, and can mask or overcompensate the
effect of the interaction on the density profile, ground-state energy, and 
other quantities. For the {\it density profile}, this means that attempts
to model the microscopically inhomogeneous charge distribution, seen 
experimentally,\cite{inhom1,inhom2,inhom3,inhom4,inhom5,inhom6,inhom7,inhom8} 
by Hubbard models that are 
homogeneous or that modulate only the interaction $U_i$, cannot lead to 
conclusive results. The influence of modulations in $v_i$ on the {\it 
ground-state energy}, on the other hand, implies that an analysis of the 
relative energetic stability of the various phases appearing in 
strongly-correlated systems is incomplete, and potentially misleading, if 
the effects of spatial inhomogeneity in these phases are not taken into 
account. All this calls in question the common practice to employ the 
homogeneous Hubbard model to model spatially inhomogeneous many-body systems, 
and demands a reconsideration of the role of nanoscale spatial inhomogeneity 
in strongly correlated systems.\cite{inhom1,inhom2,inhom3,inhom4,inhom5,inhom6,inhom7,inhom8,superl1,superl2,superl3,superl4,superl5,superl6,superl7,superl8,superl9,superl10}

{\bf Acknowledgments}\\
This work was sup\-por\-ted by FAPESP and CNPq.

\end{document}